\documentstyle[11pt,newpasp,twoside,epsf]{article} 
\def\edcomment#1{\iffalse\marginpar{\raggedright\sl#1\/}\else\relax\fi} 
\reversemarginpar 

\begin{document} 
\title{Rotation of Hot Horizontal Branch Stars in Galactic Globular Clusters.}

\author{Alejandra Recio-Blanco} 
\affil{Istituto Nazionale di Astrofisica, 
vicolo dell'Osservatorio 2 Padova, I-35122, ITALY}
\author{Giampaolo Piotto} 
\affil{Dipartimento di Astronomia, Universit\`a. di Padova, 
vicolo dell'Osservatorio 2 Padova, I-35122, ITALY} 
\author{Antonio Aparicio} 
\affil{IAC, Via Lactea s/n, 
382002 La Laguna Tenerife, SPAIN} 
\author{Alvio Renzini} 
\affil{ESO, Karl-Schwarzschild-Str. 2, D-85748 Garching bei 
M$\ddot{u}$nchen, GERMANY}  

\begin{abstract} 

We present  high resolution UVES+VLT  spectroscopic observations of 61
stars in the extended blue horizontal branches of the Galactic globular
clusters NGC~1904 (M79), NGC~2808, NGC~6093 (M80),  and NGC~7078 M15).
 Our data reveal
for the first time the presence in NGC~1904 of a sizable population of
fast ($v$sin$i\ge20$  km/s) horizontal branch (HB)  rotators, confined
to the cool  end of the  EHB, similar to  that found in M13.  We  also
confirm  the fast rotators  already  observed in NGC~7078.  The cooler
stars  (T$_{\rm eff}$ $<$   11,500 K) in these   three clusters show a
range of rotation  rates, with a group of  stars rotating at $\sim$ 15
km/s  or less, and     a fast rotating   group  at  $\sim$ 30    km/s.
Apparently, the fast rotators are relatively more abundant in NGC~1904
and M13, than  in NGC~7078.  No  fast rotators have been identified in
NGC~2808 and NGC 6093.  All the stars hotter than T$_{\rm eff}$ $\sim$
11,500 K have projected rotational velocities  $v$sin$i<$ 12 km/s. The 
connection  between
photometric gaps in the HB and the change  in the projected rotational
velocities  is not confirmed  by the new  data. However,  our data are
consistent with a relation between this discontinuity and the HB jump.

\end{abstract}

We determined the radial velocity (v$_{rad}$) and the projected 
rotational velocity ($v$sin$i$) for each of our program stars using 
the cross-correlation technique described by Tonry \& Davis 
(1979).
Figure 1 shows the complete set of projected rotational velocity data 
for M79, NGC2808, M80 and M15, including Behr et a.\ (2000b, B00b) results
for the last cluster.
These results suggest a link between the photometric G99 jump and the
discontinuity in the distribution of the HB stellar rotation rates.
Radiative levitation of metals has been invoked by G99 to explain the jump.
Besides, the metal abundance anomalies (enhancements of metals,
underabundance of He) found by Behr et al (1999) and B00b constitute an
empirical evidence that radiative levitation and diffusion are
effectively at work in the envelope of HB stars with $T_{\rm eff}\ge
11,500$K. As a consequence, R02 argue that  the
absence of fast rotators at $T_{\rm eff}\ge 11,500$K might be due to an increase of 
angular momentum removal caused by enhanced mass loss in the more 
metallic atmospheres of the stars hotter than the jump (see also 
Vink and Cassisi, 2002).  
However, the real open problem is that we still lack of any plausible 
explanation for the presence of fast rotators (see discussion in 
R02).

\begin{figure}
\vspace{-.1in}
\plottwo{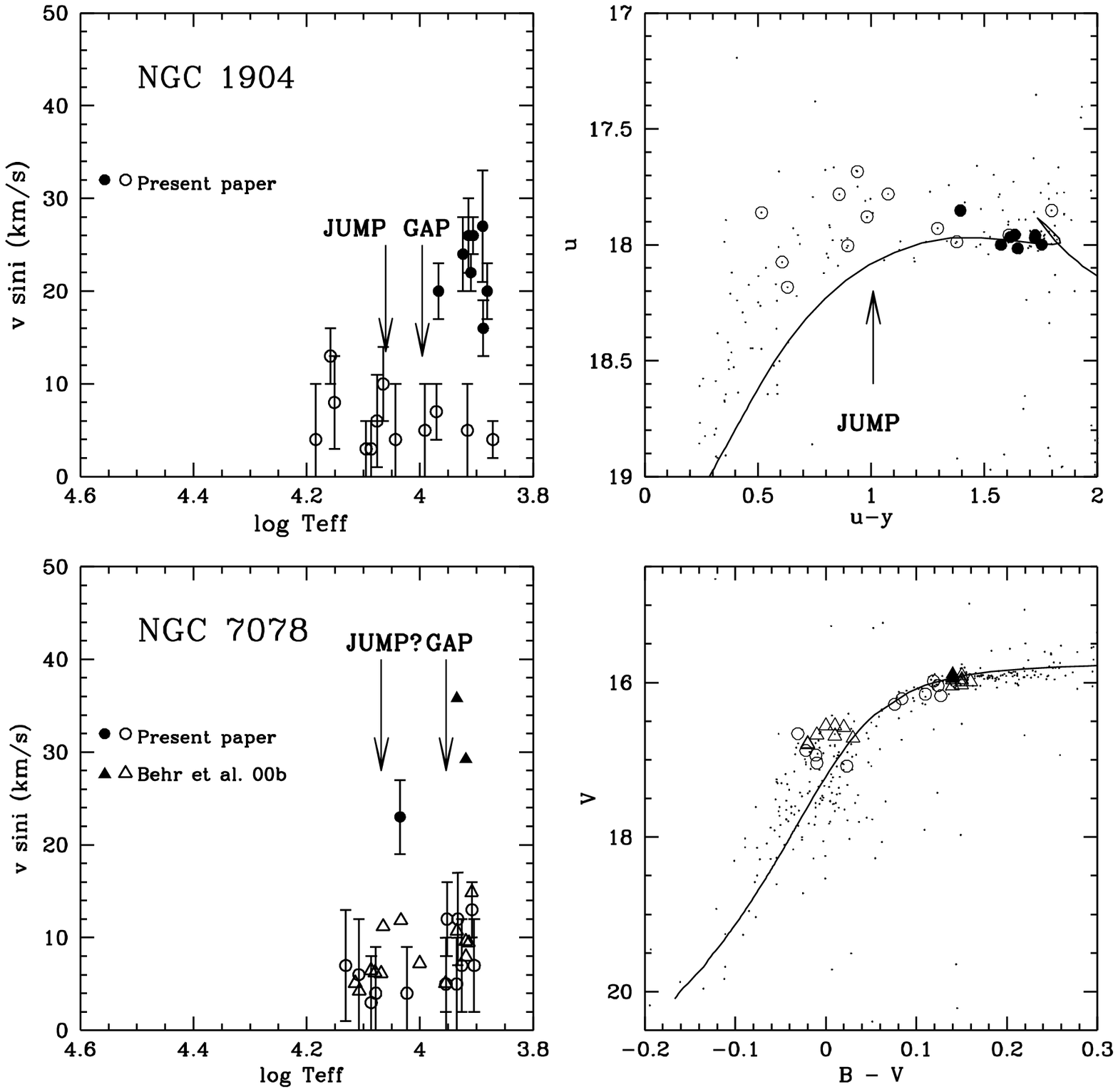}{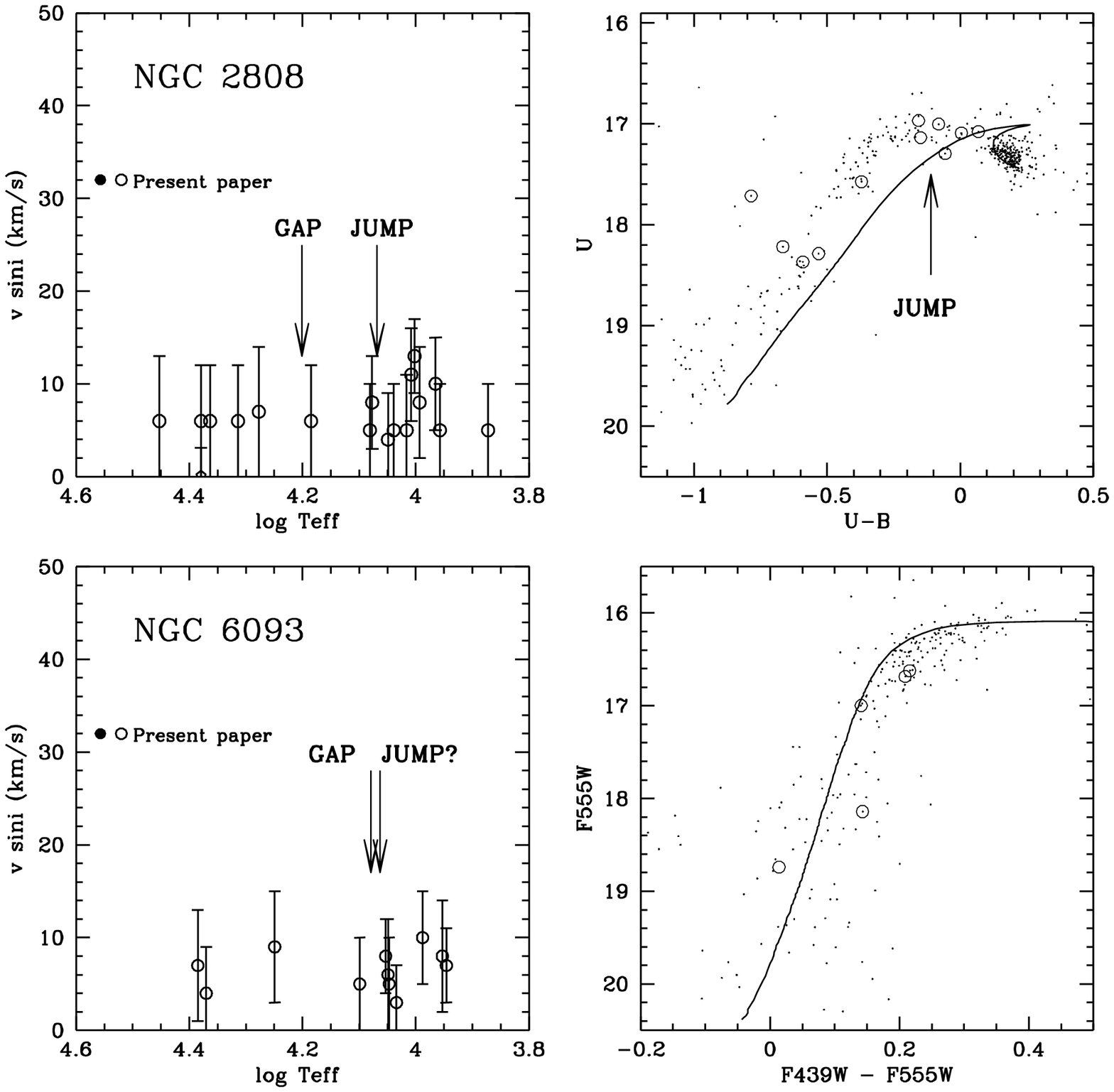}
\caption{Projected rotational
velocities as a function of  the temperature for our targets (circles)
plus  Behr et al.\ (2000)  results for  M15 (triangles).  Full symbols
show  $v$sin$i\ge15$   km/s stars.The   vertical arrows   indicate the
positions of the  ``gaps'' (from Ferraro et al.\   1998 and Piotto  et
al.\ 1999) and of  the ``jump''  from  G99.  The
position for the  jump in M80  has not been confirmed by observations,
yet. We just put the arrow  in correspondence of $T_{\rm eff}=11,500$K
where  all   the  HBs of  G99   GCs show  this
feature.  The   location of the   target stars  in the   CMDs 
(by Bedin et al. 2000, G99, and Piotto et al. 2002) is also
marked. The full line represent the best  fitting ZAHB from Cassisi et
al.\ (1999). }
\end{figure} 
%


\begin{references}

\reference Behr, B.B., Cohen, J. G., McCarthy, J.\ K., \& Djorgovski, S.\ G. 1999, 
ApJ, 517,L31 

\reference Behr, B.B.,Djorgovski, S. G., Cohen, J. G., McCarthy, J.\ K., C$\hat{o}$t$\acute{e}$, P.,  
Piotto, G., \& Zoccali, M. 2000, ApJ, 528, 849

\reference Behr, B.B., Cohen, J. G., \& McCarthy, J.\ K. 2000b, ApJ, 531,L37 

\reference Grundahl, F., Catelan, M., Landsman, W. B., Stetson, P. B., \& Andersen, 
M.\  I. 1999 ApJ , 524, 242 

\reference Recio-Blanco, A., Piotto, G., Aparicio, A. \& Renzini, A. 2002, ApJL, 572, 71


\end{references}
\end{document}